\documentclass{elsart}

\usepackage{psfrag}
\usepackage{times}
\usepackage{amsmath}
\usepackage{graphicx}

\begin{document}
\begin{frontmatter}
  
  \title{Geometrical vs. Fortuin-Kasteleyn Clusters \\ in the \\ 
    Two-Dimensional $q$-State Potts Model}
\author{Wolfhard Janke and Adriaan M. J.  Schakel}
\address{Institut f\"ur Theoretische Physik, Universit\"at Leipzig,
Augustusplatz 10/11, 04109 Leipzig, Germany }

\begin{abstract}
  The tricritical behavior of the two-dimensional $q$-state Potts model
  with vacancies for $0\leq q \leq4$ is argued to be encoded in the
  fractal structure of the geometrical spin clusters of the pure model.
  The known close connection between the critical properties of the pure
  model and the tricritical properties of the diluted model is shown to
  be reflected in an intimate relation between Fortuin-Kasteleyn and
  geometrical clusters: The same transformation mapping the two critical
  regimes onto each other also maps the two cluster types onto each
  other.  The map conserves the central charge, so that both cluster
  types are in the same universality class.  The geometrical picture is
  supported by a Monte Carlo simulation of the high-temperature
  representation of the Ising model ($q=2$) in which closed graph
  configurations are generated by means of a Metropolis update algorithm
  involving single plaquettes.

\end{abstract}
\date{\today}
\end{frontmatter}

\section{Introduction}
The two-dimensional $q$-state Potts models \cite{Potts} can be
equivalently formulated in terms of Fortuin-Kasteleyn (FK) clusters of
like spins \cite{FK}.  These FK clusters are obtained from the
geometrical spin clusters, which consist of nearest neighbor sites with
their spin variables in the same state, by laying bonds with a certain
probability between the nearest neighbors.  The resulting FK, or bond
clusters are in general smaller than the geometrical ones and also more
loosely connected.  The FK formulation of the Potts models can be
thought of as a generalization of (uncorrelated) bond percolation, which
obtains in the limit $q\to1$.  The geometrical clusters themselves arise
in the low-temperature representation of the pure model \cite{Wu}.

For $q \le 4$, where the model undergoes a continuous phase transition,
the FK clusters percolate at the critical temperature and their fractal
structure encodes the complete critical behavior.  The thermal critical
exponents are obtained using the cluster definitions from percolation
theory \cite{StauferAharony}.  With clusters and their fractal
properties taking the central stage, the FK formulation provides a
geometrical description of the Potts model.  

The concept of correlated bond percolation has been turned into a
powerful Monte Carlo algorithm by Swendsen and Wang \cite{SwendsenWang},
and by Wolff \cite{Wolff}, in which not individual spins are updated,
but entire FK clusters.  The main advantage of the nonlocal cluster
update over a local spin update, like Metropolis or heat bath, is that
it drastically reduces the critical slowing down near the critical
point.

Although it was known from the relation with other statistical models
that the phase transition of the Potts models changes from being
continuous to first order at $q=4$ \cite{Baxter}, initial
renormalization group approaches failed to uncover the first-order
nature for larger $q$.  Only after the pure model was extended to
include vacant sites, this feature was observed \cite{NBRS}.  In a
Kadanoff block-spin transformation, the vacant sites represent blocks
without a majority of spins in a certain state, i.e., they represent
disordered blocks.  In addition to the pure Potts critical behavior, the
site diluted model also displays tricritical behavior, which was found
to be intimately connected with the critical behavior \cite{NBRS}.  With
increasing $q$, the critical and tricritical fixed points move together
until at $q=4$ they coalesce and the continuous phase transition turns
into a first-order one.

\begin{figure}
\begin{center}
\includegraphics[width=3.0cm]{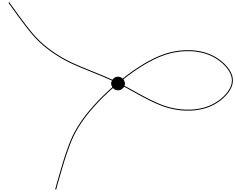}
\end{center}
\caption{Trace possessing a double point.
\label{fig:double}}
\end{figure}

Recently, the cluster boundaries of two-dimensional critical systems
have been intensively studied by means of a method dubbed ``stochastic
Loewner evolution'' (SLE$_{\bar\kappa}$)--a one-parameter family of
random conformal maps, introduced by Schramm \cite{Schramm}.  In this
approach, the Brownian motion of a random walker is described by
Loewner's ordinary differential equation, containing a random term whose
strength is specified by a parameter $\bar\kappa \ge 0$ \cite{footnote}.
Different values of $\bar\kappa$ define different universality classes.
Various results previously conjectured on the basis of the Coulomb gas
map \cite{denNijs,Nienhuis,SD} and conformal invariance \cite{Cardyrev}
have been rigorously established by this method.  The interrelation
between SLE$_{\bar\kappa}$ traces and the Coulomb gas description is
made explicit by observing that the Coulomb gas coupling parameter $g$
can be simply expressed by $g=1/\bar{\kappa}$ \cite{Duplantier02}.  The
nature of the SLE$_{\bar\kappa}$ traces changes with $\bar\kappa$: for
$0 \le \bar\kappa < 1$ the path is simple (non-intersecting), while for
$1 \leq \bar\kappa \leq 2$ it possesses double points (see
Fig.~\ref{fig:double}); and for $\bar\kappa > 2$ it is space-filling
\cite{RS,BB}.
This change reflects a change in the critical systems described by the
SLE$_{\bar\kappa}$ traces \cite{Duplantier02}: For $1 \le \bar\kappa \le
2$ they represent the hulls of FK clusters of the $q$-state Potts model
with $4 \ge q \ge 0$, while for $\frac{1}{2} \le \bar\kappa \le 1$ they
represent the closed graphs of the high-temperature representation of
the O($n$) model with $-2 \le n \le 2$ and at the same time also the
external perimeters of FK clusters with dual parameter $1/\bar\kappa$
\cite{Duplantier02,Duplantiermath}.  For a recent overview from
the mathematical point of view, see Ref.~\cite{Duplantiermath}.

In this paper, the tricritical regime of the two-dimensional (annealed)
site diluted $q$-state Potts model with $0\le q \le4$ is studied from
the geometrical point of view.  It will be argued that the tricritical
behavior of these models is encoded in the geometrical clusters of the
pure Potts model in the same way that the critical behavior is encoded
in the FK clusters.  The relation between geometrical clusters and
tricritical behavior was first established by Stella and Vanderzande
\cite{StellaVdzandePRL,VdzandeStellaJP} for the special case $q=2$,
i.e., for the Ising model.  Using arguments based on renormalization
group, conformal invariance, and numerical simulations, they showed that
the geometrical cluster dimensions of the Ising model at criticality are
determined by the $q=1$ tricritical Potts model, as was earlier
conjectured by Temesv\'ari and Her\'enyi \cite{TeHe}.  The values of two
of the three leading tricritical exponents characterizing the
geometrical clusters were already determined before by Coniglio and
Klein \cite{ConKl}.  The distinctive feature of the $q=1$ tricritical
model is that it is in the same universality class as the Ising model
defined by the central charge $c=\frac{1}{2}$. In addition, it has the
same correlation length exponent $\nu=1$ as the Ising model.  The
boundaries of geometrical Potts clusters were also already known to be
in the same universality class as the tricritical model with the same
central charge \cite{Duplantier87,SD}.  Formulated in terms of
SLE$_{\bar\kappa}$ traces, the hulls of geometrical clusters are
described at criticality by traces with $\frac{1}{2} \le \bar\kappa \le
1$, thereby forming the geometrical counterpart of the FK hulls which
are described by the traces with $1 \le \bar\kappa \le 2$.  It will be
shown here that not just the boundary dimensions, but all the relevant
fractal dimensions characterizing FK clusters are in one-to-one
correspondence with those of the geometrical clusters, thus providing a
physical picture for the close connection between the critical and
tricritical behaviors just mentioned.  The important aspect of the map
is that it leaves the central charge unchanged, so that both cluster
types and their boundaries are in the same universality class,
characterized by the same central charge.

To support this geometrical picture, we carry out a Monte Carlo study of
the high-temperature representation of the 2-state Potts, or Ising
model.  We generate the high-temperature graphs by using a Metropolis
update algorithm involving single plaquettes \cite{Erkinger}.  By
duality, the high-temperature graphs, which are closed, form the hulls
of geometrical spin clusters on the dual lattice.  We thus simulate the
geometrical hulls directly without first considering the clusters.  From
the geometrical properties of these graphs, such as their distribution,
the size of the largest graph, and whether or not a graph spans the
lattice, the fractal dimension of the hulls of geometrical clusters can
be determined immediately.  Our numerical result agrees with the
analytic prediction by Duplantier and Saleur \cite{DS88}, which was
derived using the Coulomb gas map.

The paper is organized as follows.  In Sec.~\ref{sec:dil}, those aspects
of the diluted $q$-state Potts model are reviewed that are of relevance
for the following, in particular its cluster properties.  Section
\ref{sec:FK} discusses the various fractal dimensions characterizing FK
clusters.  In Sec.~\ref{sec:G}, the results for FK clusters are
transcribed to geometrical clusters, which in Sec.~\ref{sec:tri} are
shown to encode the tricritical Potts behavior.  The Monte Carlo results
for the high-temperature representation of the Ising model are presented
in Sec.~\ref{sec:HT}, followed by a summary in Sec.~\ref{sec:summary}.

\section{Diluted $q$-state Potts Model}
\label{sec:dil}
The diluted Potts model can be defined by the Hamiltonian \cite{Murata}:
\begin{equation} 
\label{delPotts}
- \beta \mathcal{H} =  K\sum_{\langle ij\rangle}
(\delta_{\sigma_i,\sigma_j} -1)
+ J \sum_{\langle ij\rangle} (\delta_{\tau_i,\tau_j}-1) 
\delta_{\sigma_i,\sigma_j} - H \sum_i (\delta_{\tau_i,1}-1), 
\end{equation} 
where $\beta$ denotes the inverse temperature and the double sum
$\sum_{\langle ij\rangle}$ extends over nearest neighbors only.  The
first term at the right hand with coupling constant $K$ is the pure
$q$-state Potts model with spin variable $\sigma_i=1,2,\cdots,q$ at the
$i$th site.  On this a second Potts model with auxiliary spin variable
$\tau_i=1,2,\cdots,s$, coupling constant $J$, and ghost field $H$ is
superimposed.  The reasons for this extension of the pure Potts model
are twofold.  First, in the limit $s\to1$, the $s$-state Potts model
describes (bond) percolation \cite{FK}, which is naturally formulated in
terms of clusters.  The extension thus allows for the investigation of
cluster properties of the original model when the limit $s\to1$ is
taken.  Second, the diluted model has two fixed points for $q\le4$.  In
addition to the pure Potts critical point, it also has a tricritical
point, depending on the value of the coupling constant $J$.  Adding
vacancies therefore leads to two distinct scaling regimes, each with its
own critical exponents.  Note that the limit $s\to 1$ is subtle as
precisely for $s = 1$, the Hamiltonian (\ref{delPotts}) is independent
of $J$ and $H$, so that it reduces to the standard Potts model.

\begin{figure}
\begin{center}
\includegraphics[width=9.0cm]{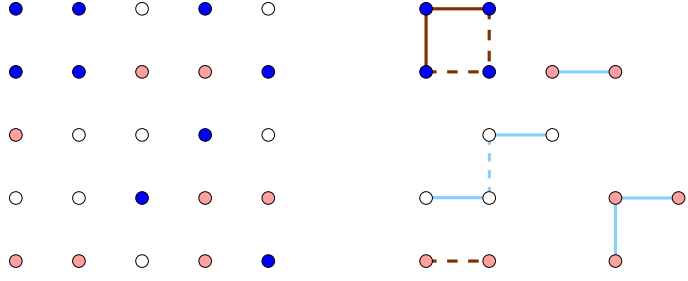}
\end{center}
\caption{{\it Left panel}: Original lattice with the spin variables
  taking three different values ($q=3$) indicated by different grey
  scales.  {\it Right panel}: Reduced lattice consisting of sites having
  at least one nearest neighbor with like spin, i.e., of the same
  shading.  Auxiliary bonds, taking two different values ($s=2$), are
  laid with a certain probability within clusters of like spins.  The
  resulting bond clusters represent sites which are all in the same spin
  (indicated by dots with the same shading) and auxiliary spin state
  (indicated by bonds with the same shading).  The broken bonds within a
  spin cluster are indicated by broken lines.
\label{fig:bond}}
\end{figure}

The second term in the Hamiltonian (\ref{delPotts}) connects nearest
neighbors of like spins (for which $\delta_{\sigma_i,\sigma_j}=1$) with
bond probability
\begin{equation} 
\label{p}
p = 1 - {\rm e}^{-J},
\end{equation}  
while the ghost field $H$ in the last term acts as a chemical potential
for the auxiliary spins in the first state, $\tau_i=1$.  A configuration of
auxiliary bonds thus obtained can be represented by a graph on a
restricted lattice, consisting of those sites of the original lattice
that have at least one nearest neighbor with like spins (see
Fig.~\ref{fig:bond}).

Following Fortuin and Kasteleyn \cite{FK}, we can rewrite the partition
function of the model (\ref{delPotts}) as \cite{Murata}
\begin{equation} 
\label{ZMurata}
Z =  \sum_{\{\sigma\}} \exp \left[ K\sum_{\langle ij\rangle}
  (\delta_{\sigma_i,\sigma_j}-1) \right]  \sum_{\{\Gamma\}}
  p^b (1-p)^{\bar{b}} \prod_{\mathcal{C}(\Gamma)} \left[1+ (s-1) {\rm e}^{-H
  n_c} \right], 
\end{equation} 
where $\{\Gamma\}$ denotes the set of bond configurations specified by
$b$ bonds and $\bar{b}$ broken bonds between like spins, while
$\mathcal{C}(\Gamma)$ denotes the clusters in a given bond configuration
$\Gamma$.  Finally, $n_c$ is the number of sites contained in the $c$th
cluster, where an isolated single site counts as a cluster, irrespective
of whether it is on the restricted or the original lattice.  This last
observation follows from the absence of any reference to the spin
variable in the last term of the Hamiltonian (\ref{delPotts}).  In the
partition function, the limiting case $s = 1$, where the Hamiltonian
(\ref{delPotts}) is independent of $J$ and $H$, can be recovered by
noting that, for a given spin configuration the sum $\sum_{\{\Gamma\}}
p^b (1-p)^{\bar{b}}$ of the probabilities of all possible bond
configurations adds up to unity.

As mentioned above, cluster properties can be extracted from the
partition function (\ref{ZMurata}) of the diluted model by taking the
limit $s\to1$.  Specifically \cite{Murata},
\begin{equation} 
\label{clustergen} 
\left. \frac{1}{N} \frac{{\rm d} \ln Z}{{\rm d}s} \right|_{s=1} =
\frac{1}{N} \left\langle 
\sum_{\{\Gamma\}} p^b (1-p)^{\bar{b}} \prod_{
\mathcal{C}(\Gamma)} {\rm e}^{-H n_c} \right\rangle = 
\sum_n \ell_n {\rm e}^{-H n},
\end{equation} 
where $N$ denotes the total number of lattice sites and $\ell_n$ is the
cluster distribution giving the average number density of clusters of
$n$ sites.  The thermal average indicated by angle brackets in
Eq.~(\ref{clustergen}) is taken with respect to the pure $q$-state Potts
model, i.e., the first factor in the Hamiltonian (\ref{delPotts}). The
right hand is seen to be the generating function for clusters.  By
differentiating it with respect to the ghost field $H$ higher momenta in
the cluster sizes can be obtained.

The pure Potts part of the theory is easily dealt with by noting that,
for a given spin configuration, it gives a factor $\left({\rm
e}^{-K}\right)^a$, where $a$ denotes the number of nearest neighbor
pairs of unlike spin.

\begin{figure}
\begin{center}
  \psfrag{K}[t][t][.8][0]{$K$} 
  \psfrag{kc}[t][t][.8][0]{$K_c$}
  \psfrag{p}[t][t][.8][0]{$p$}
  \psfrag{Geo}[l][t][.8][0]{Geometrical}
  \psfrag{FK}[l][t][.8][0]{FK}
  \includegraphics[width=3.0cm]{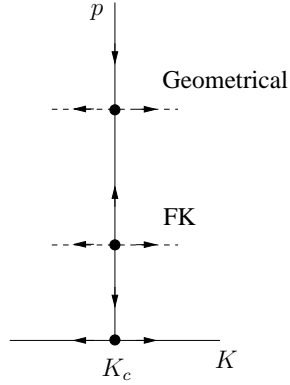}
\end{center}
\caption{Phase diagram of the diluted model (\ref{delPotts}) in the
  $K$-$p$ plane, with $p=1-\exp(-J)$ the bond probability.  The arrows
  indicate the renormalization flow in the infrared.  The FK fixed point
  governs the critical behavior described by the FK clusters, while the
  geometrical fixed point governs the tricritical behavior described by
  the geometrical clusters.
  \label{fig:phasediagram} }
\end{figure}

The two fixed points of the diluted model correspond to two specific
choices of the coupling constant $J$ \cite{ConKl,Vanderzande} with $K$
fixed at the critical temperature $K_\mathrm{c}$ of the pure Potts model
(see Fig.~\ref{fig:phasediagram}).
The first choice is obtained by taking $J=K$.  This case is special
because the factors ${\rm e}^{-K}$ arising from the first term in the
Hamiltonian (\ref{delPotts}) can now be related to the bond probability
(\ref{p}) as
\begin{equation} 
{\rm e}^{-K} = 1 -p,
\end{equation} 
and the partition function becomes
\begin{equation} 
Z_{\rm FK} = \sum_{\{\Gamma\}} p^b (1-p)^{B-b} \prod_{\mathcal{C}(\Gamma)}
  \left[1+ (s-1) {\rm e}^{-H n_c } \right] q,
\end{equation} 
where $B$ is the total number of bonds on the lattice, $B= b + \bar{b} +
a$.  The sum $\sum_{\{\sigma\}}$ produced the factor $q$ since each
cluster can be in any of the $q$ spin states.  For $s = 1$, this
partition function reduces to the celebrated Fortuin-Kasteleyn
representation of the Potts model \cite{FK},
\begin{equation} 
\label{ZFK}
Z_{\rm FK} =  \sum_{\{\Gamma\}} p^b (1-p)^{B - b} q^{N_\mathrm{C}},
\end{equation} 
where $N_\mathrm{C}$ is the number of clusters, including isolated
sites, contained in the bond configuration $\Gamma$.  The clusters seen
in the limit $s\to1$ encode the complete thermal critical behavior of
the model and are frequently referred to as Fortuin-Kasteleyn (FK)
clusters.  For $q \to 1$, the partition function (\ref{ZFK}) describes
standard, uncorrelated percolation, where the FK clusters coincide with
the usual percolation clusters.

The second choice is obtained by taking the limit $J\to \infty$ where
the bond probability $p$ tends to unity.  The only clusters surviving
this limit are those without any broken bonds between like spins
($\bar{b}=0$), which are the geometrical clusters.  The partition
function can be written in this limit as
\begin{equation} 
\label{Zgeogen}
Z_{\rm G} = \sum_{\{\Gamma\}} {\rm e}^{-K a}
  \prod_{\mathcal{C}(\Gamma)} \left[1+ (s-1) {\rm e}^{-H n_c } \right]
   P_c(q),
\end{equation} 
where the factor $P_c(q)$ is such that the product over the clusters
$\prod_{\mathcal{C}(\Gamma)} P_c(q) =: P_\Gamma(q)$ gives the number of
different spin configurations for a given bond configuration $\Gamma$.
That is, $P_\Gamma(q)$ is the number of $q$-colorings of the geometrical
clusters contained in $\Gamma$, where it is recalled that an isolated
single site counts as a cluster.  For $s= 1$, the partition function
reduces to
\begin{equation} 
\label{Zgeo}
Z_{\rm G} = \sum_{\{\Gamma\}} {\rm e}^{-K a} P_\Gamma(q),
\end{equation} 
which is nothing but the standard low-temperature representation of the
pure Potts model \cite{Wu}.  For the Ising model ($q=2$), each graph can
be colored in two different ways, $P_\Gamma(2)=2$, so that the coloring
number becomes irrelevant.  For uncorrelated percolation ($q=1$),
$P_\Gamma(1)=1$ and $a=0$ trivially, so that only one geometrical
cluster remains, representing a fully occupied lattice
\cite{Vanderzande}.

\section{Pure Potts Model}
\subsection{FK Clusters}
\label{sec:FK}
Adapting similar notations \cite{Nienhuis}, we parameterize the
two-dimensional $q$-state Potts models as
\begin{equation}
\label{Potts_branch} 
\sqrt{q} = - 2 \cos(\pi/\bar\kappa), 
\end{equation} 
with $2 \ge \bar\kappa \ge 1$ so that the argument of the cosine takes
values in the interval $[\pi/2 ,\pi]$.  Special cases are:
\begin{itemize}
\item tree percolation ($q=0, \; \bar\kappa = 2$) 
\item uncorrelated percolation ($q=1, \; \bar\kappa = \frac{3}{2}$)
\item Ising model ($q=2, \; \bar\kappa = \frac{4}{3}$)
\item $q=3, \; \bar\kappa = \frac{6}{5}$
\item $q=4, \; \bar\kappa = 1$.
\end{itemize}
The parameter $\bar\kappa$ is related to the central charge $c$ via
\cite{Cardyrev}
\begin{equation} 
\label{conf}
c = 1 - \frac{6(1-\bar\kappa)^2}{\bar\kappa},
\end{equation} 
while the correlation length exponent $\nu$ and the Fisher exponent
$\eta_\mathrm{C}$ are given by \cite{denNijs}:
\begin{equation}
\label{cesPotts} 
\nu = \frac{2}{3} \frac{1}{2-\bar\kappa}, \quad \eta_\mathrm{C} = 2 -
\frac{1}{\bar\kappa} - \frac{3}{4} \bar\kappa.
\end{equation}
The latter determines the algebraic decay of the cluster correlation
function $G_\mathrm{C}({\bf x})$ at the critical point:
\begin{equation} 
G_\mathrm{C}({\bf x}, {\bf x}') \sim 1/|{\bf x}-{\bf x}'|^{d-2 +
  \eta_\mathrm{C}}, 
\end{equation} 
with $d$ the number of space dimensions. Physically, $G_\mathrm{C}({\bf
  x}, {\bf x}')$ gives the probability that sites ${\bf x}$ and
${\bf x}'$ belong to the same cluster.  The subscript ``C'' is to
distinguish the thus defined exponent from the standard definition based
on the spin-spin correlation function.  The other exponents can be
obtained from the two given in Eq.~(\ref{cesPotts}) using standard
scaling relations.

For ease of comparison with the more often used notation where the
central charge is given in terms of a parameter $m$ \cite{Cardyrev}:
\begin{equation} 
\label{cm}
c = 1 - \frac{6}{m(m+1)},
\end{equation} 
which in turn is related to $q$ through
\begin{equation}
\label{qmPotts} 
\sqrt{q} = 2 \cos \left(\frac{\pi}{1+m}\right) = -2  \cos \left(
  \frac{m}{1+m} \pi\right),  
\end{equation}  
with $1\leq m \leq \infty$, we note that the relation with $\bar\kappa$
reads for Potts models:
\begin{equation} 
\label{kappam}
\bar\kappa =  \frac{1+m}{m}, \quad m = \frac{1}{\bar\kappa -1}.
\end{equation} 
Usually, only the first equation in Eq.~(\ref{qmPotts}) is given, we
included the second to clearly see the relation with the $\bar\kappa$
parameterization.

The critical behavior of the Potts model is also encoded in the FK cluster
distribution $\ell_n$ given in Eq.~(\ref{clustergen}), which near
the critical point takes the form
\begin{equation} 
\ell_n \sim n^{- \tau_{\rm C}} \exp(- \theta n),
\end{equation} 
as in percolation theory \cite{StauferAharony}.  The first factor,
characterized by the exponent $\tau_{\rm C}$, is an entropy factor,
measuring the number of ways of implementing a cluster of given size on
the lattice.  The second factor is a Boltzmann weight which suppresses
large clusters as long as the parameter $\theta$ is finite.  When the
critical temperature is approached from above, it vanishes as $\theta
\propto (T-T_{\rm c})^{1/\sigma_{\rm C}}$, with $\sigma_{\rm C}$ a
second exponent.  The cluster distribution then becomes algebraic,
meaning that clusters of all sizes are present.  As in percolation
theory \cite{StauferAharony}, the values of the two exponents specifying
the cluster distribution uniquely determine the critical exponents
as
\begin{align} 
\label{perce}
\alpha & = 2 - \frac{\tau_{\rm C} -1}{\sigma_{\rm C}}, & \beta_{\rm C}
&= \frac{\tau_{\rm C} -2}{\sigma_{\rm C}}, & \gamma_{\rm C} & =
\frac{3-\tau_{\rm C}}{\sigma_{\rm C}}, \nonumber \\ \eta_{\rm C} &= 2 +
d \frac{\tau_{\rm C}-3}{\tau_{\rm C}-1}, & \nu & = \frac{\tau_{\rm C}
-1}{d \sigma_{\rm C}}, & D_\mathrm{C} & = \frac{d}{\tau_{\rm C}-1},
\end{align} 
where the fractal dimension of the clusters is related to the Fisher
exponent (\ref{cesPotts}) via
\begin{equation} 
\label{Deta}
D_\mathrm{C} = \tfrac{1}{2} (d+2-\eta_{\rm C}).
\end{equation} 
Various exponents are given the subscript ``C'' to indicate that the
cluster definition is used in defining them.  For FK clusters, where in
terms of $\bar\kappa$
\begin{equation} 
\label{sigmatau}
\sigma_{\rm C} = \frac{12 \bar\kappa (2 - \bar\kappa)}{3 \bar\kappa^2 +
  8 \bar\kappa +4}, \quad \tau_{\rm C} = \frac{3 \bar\kappa^2 + 24
  \bar\kappa +4}{3 \bar\kappa^2 + 8 \bar\kappa +4},
\end{equation} 
the cluster exponents coincide with the thermal ones.  The cluster
definition not always yields the thermal critical exponents.  For
example, when the critical behavior of a system allows for a description
in terms of other geometrical objects such as closed particle worldlines
or vortex loops, a related but different definition is required to
obtain the thermal exponents from the loop distribution
\cite{vortexland}.

The various fractal dimensions characterizing FK clusters and the
leading thermal eigenvalue $y_T=1/\nu$ read in terms of $\bar\kappa$
\cite{SD,Coniglio1989,Duplantier00}
\begin{subequations}
\label{DPotts}
\begin{eqnarray} 
D_\mathrm{C}  &=& 1 + \frac{1}{2\bar\kappa} + \frac{3}{8} \bar\kappa \\
D_\mathrm{H} &=& 1 + \frac{\bar\kappa}{2}  \\
D_\mathrm{EP} &=& 1 + \frac{1}{2\bar\kappa}  \\
D_\mathrm{RB} &=& 1 - \frac{3}{2\bar\kappa} + \frac{\bar\kappa}{2} \\
y_T &=& 3 -  \frac{3}{2} \bar\kappa .
\end{eqnarray} 
\end{subequations}
Here, $D_\mathrm{C}$ is the fractal dimension of the clusters
themselves, and $D_{\rm H}$ that of their hulls \cite{GrossmanAharony,ABA}.
In the context of uncorrelated percolation, the hull of a cluster can be
defined as a biased random walk \cite{GrossmanAharony}: Identify two
endpoints on a given cluster.  Starting at the lower endpoint, the
walker first attempts to move to the nearest neighbor to its left.  If
that site is vacant the walker attempts to move straight ahead.  If that
site is also vacant, the walker attempts to move to its right.  Finally,
if also that site is vacant, the walker returns to the previous site,
discards the direction it already explored and investigates the (at most
two) remaining directions in the same order.  When turning left or
right, the walker changes its orientation accordingly.  The procedure is
repeated iteratively until the upper endpoint is reached.  To obtain the
other half of the hull, the entire algorithm is repeated for a random
walker that attempts to first move to its right instead of to its left.
The hull of FK clusters is a self-intersecting path.  As remarked in the
Introduction, these hulls for $4 \ge q \ge 0$ correspond to the
SLE$_{\bar\kappa}$ traces with $1 \le \bar\kappa \le 2$.

The fractal dimension $D_{\rm EP}$ characterizes the external perimeter.
Its operational definition \cite{GrossmanAharony} in the context of
uncorrelated percolation is similar to that for the hull with the
proviso that the random walker only visits (nearest neighbor) vacant
sites around the hull of the cluster.  From the resulting trace those
sites not belonging to the perimeter, i.e., without an occupied site as
nearest neighbor, are deleted (such sites can, for example, be visited
by the walker when it makes a right or left turn on a square lattice).
This leads to non-intersecting traces.

These algorithms have recently been used in a numerical study carried
out on fairly large lattices ($L=2^{12}=4096$) to determine the fractal
dimensions of FK clusters of the $q$-state Potts models with $q=1,2,3,4$
\cite{AAMRH}.

By construction, the external perimeter is smoother than the hull, or at
least as smooth, so that $D_{\rm EP} \leq D_{H}$.  The two are equal
when the fractal dimension $D_{\rm RB}$ of the so-called red bonds
\cite{Stanley} is negative.  (A red bond denotes a bond that upon
cutting leads to a splitting of the cluster.)  For FK clusters, $D_{\rm
  EP}$ is strictly smaller than $D_{\rm H}$ for all $1 < \bar\kappa \le
2$, while for $\bar\kappa=1$ ($q=4$), the fractal dimension of the red
bonds becomes zero, and the hull and the external perimeter have
identical dimensions.

The two boundary dimensions are seen to satisfy the relation
\cite{Duplantier00}
\begin{equation} 
\label{Drel}
(D_\mathrm{EP}-1) (D_\mathrm{H}-1) = \tfrac{1}{4}.
\end{equation}
An analogous relation, this time also involving the fractal dimension of
the FK clusters themselves reads
\begin{equation}
\label{Da} 
(D_\mathrm{C} - D_\mathrm{EP}) (D_\mathrm{C} - \tfrac{3}{4} D_\mathrm{H}
- \tfrac{1}{4}) = \tfrac{3}{16}.
\end{equation} 
In addition, the fractal dimensions satisfy the linear relation
\begin{equation} 
D_\mathrm{C} - D_\mathrm{H} = \tfrac{1}{4} (D_\mathrm{EP} -
D_\mathrm{RB}).
\end{equation} 
The dimension $D_\mathrm{C}$ of the FK clusters approaches the number of
available dimensions, $D_{\rm C} \to 2$ when $q\to0$ ($\bar\kappa \to
2$).  In this limit, also the hulls of the FK clusters become
space-filling, $D_{\rm H} \to 2$.

As in percolation theory \cite{Stanley}, the fractal dimensions can be
identified with renormalization group eigenvalues of certain operators.
For example, the fractal dimension $D_\mathrm{C}$ of the FK clusters
coincides with the magnetic scaling exponent $y_H$,
\begin{equation}  
\label{yH}
D_\mathrm{C} = y_H = d - \beta_\mathrm{C}/\nu,
\end{equation} 
while that of the red bonds, $D_{\rm RB}$, coincides with the eigenvalue
$y_J$ in the $J$ direction,
\begin{equation}  
D_{\rm RB} = y_J,
\end{equation} 
and therefore describes the crossover between FK and geometrical
clusters \cite{StellaVdzandePRL,Coniglio1989}.  Specifically, $1/y_J$
determines the divergence of the correlation length $\xi$ at the
critical temperature $K_\mathrm{c}$ when the bond probability (\ref{p})
approaches the critical value $p_\mathrm{c}= 1 - \exp(-J_\mathrm{c})$,
i.e., $\xi \sim (p_\mathrm{c} - p)^{-1/y_J}$.

\subsection{Geometrical Clusters}
\label{sec:G}
Starting from the FK cluster dimensions (\ref{DPotts}), we next wish to
obtain the analog expressions for the geometrical clusters of the Potts
model at criticality.  As argued by Vanderzande \cite{Vanderzande}, both
cluster types are characterized by the same central charge $c$.  From
Eq.~(\ref{conf}) it follows that a given value of $c$ does not uniquely
determine $\bar\kappa$.  Indeed, inverting that equation, we obtain {\it
  two} solutions for $\bar\kappa$:
\begin{equation} 
\label{kappac}
\bar\kappa_\pm = \frac{13-c \pm \sqrt{(c-25)(c-1)}}{12},
\end{equation}
with $\bar\kappa_+\ge1$ and $\bar\kappa_-\le1$.  The solutions satisfy
the constraint
\begin{equation}
\label{constraint}
\bar\kappa_+ \bar\kappa_- = 1.
\end{equation} 
Since the right hand is independent of $c$, replacing $\bar\kappa$ with
$1/\bar\kappa$ leaves the central charge unchanged, $c({\bar\kappa}) =
c(1/{\bar\kappa})$.  

As an aside, note that the fractal dimensions of the hull and external
perimeter of FK clusters are related by precisely this map
\cite{Duplantier00}, which is called duality in SLE studies.  This
duality, which in the Coulomb gas language corresponds to the earlier
observed correspondance $g \to 1/g$ (recall that $g=1/\bar{\kappa}$)
\cite{Duplantier02,Duplantiermath}, is also at the root of the relation
(\ref{Drel}) between the two boundary dimensions.

Applying the duality map $\bar\kappa \to 1/\bar\kappa$ to the FK cluster
dimensions listed in Eq.~(\ref{DPotts}), we obtain
\begin{subequations}
\label{Dgeo}
\begin{eqnarray} 
\label{DgeoD}
D_\mathrm{C}^{\rm G} &=& 1 + \frac{3}{8\bar\kappa} + \frac{\bar\kappa}{2} 
\label{DGC} \\
D^{\rm G}_\mathrm{H} &=& 1 + \frac{1}{2 \bar\kappa} \label{DGH} \\
D^{\rm G}_\mathrm{RB} &=& 1 + \frac{1}{2\bar\kappa} - \frac{3}{2}
\bar\kappa \label{DRBgeo} \\ 
y^{\rm G}_T &=& 3 - \frac{3}{2\bar\kappa}, \label{yGT}
\end{eqnarray} 
\end{subequations}
where $\bar\kappa = \bar\kappa_+$.  These dimensions precisely match
those conjectured by Vanderzande for geometrical clusters
\cite{Vanderzande}.  We therefore conclude that the geometrical clusters
of the Potts model are images of the FK clusters under the map
$\bar\kappa \to 1/\bar\kappa$ for given $\bar\kappa$.  Since this
transformation at the same time maps the critical onto the tricritical
regime (see below), it follows that the geometrical clusters (of the
pure Potts model) describe the tricritical behavior in the same way as
the FK clusters describe the critical behavior.  For $\bar\kappa=1$
($q=4$), the dimensions of the FK and geometrical clusters become
degenerate, and the critical and tricritical behaviors merge.  For
$q>4$, the phase transition is discontinuous \cite{Baxter}.  The fractal
dimension (\ref{DGC}) was first given by Stella and Vanderzande
\cite{StellaVdzandePRL} for the Ising case, and generalized to arbitrary
$1 \le \bar\kappa \le 2$ by Duplantier and Saleur \cite{DS89}.

The external perimeter dimension $D_{\rm EP}$ of FK clusters has no
image under the map $\bar\kappa \to 1/\bar\kappa$.  To understand this,
recall that by smoothing the hull of a FK cluster one obtains its
external perimeter.  For geometrical clusters, on the other hand, the
fractal dimension of the red bonds is negative $D_{\rm RB} \le 0$ as
follows from Eq.~(\ref{DRBgeo}) with $1\le \bar\kappa \le 2$, so that
$D^{\rm G}_{\rm H}= D^{\rm G}_{\rm EP}$ and the hull of a geometrical
cluster is already non-intersecting.  This is also reflected by the
SLE$_{\bar\kappa}$ traces.  Under the transformation $\bar\kappa \to
1/\bar\kappa$, the self-intersecting traces with $\bar\kappa \ge 1$,
representing the hulls of FK clusters are mapped onto simple traces with
$\bar\kappa \le 1$, representing the hulls of geometrical clusters
\cite{Duplantier02}.

For given $\bar\kappa$, the fractal dimension of the external perimeters
of FK clusters coincides with that of the hulls of geometrical clusters,
$D_{\rm EP} = D^{\rm G}_{\rm H} (= D^{\rm G}_{\rm EP})$.  Since FK
clusters are obtained from geometrical clusters by breaking bonds
between nearest neighbors with like spins, the smoothing of the FK hulls
apparently undoes this process again (as far as the boundaries are
concerned).

Table~\ref{table:Ds} summarizes the various fractal dimensions appearing
in the Potts models. Note that for $q=0$, the geometrical cluster
dimension is larger than the number of available dimensions, $D_{\rm
  C}^{\rm G}>2$, making the geometrical clusters unphysical in this
case.  The equality $D_{\rm EP} = D_{\rm RB}$ is typical for tree
percolation.  The value $D^{\rm G}_{\rm C}=2$ for $q=1$ agrees with the
observation that in this case the geometrical cluster represents a fully
occupied lattice \cite{Vanderzande}.  The cluster dimension $D_{\rm
  C}(\bar\kappa)$ possesses a minimum at $1< \bar\kappa=2/\sqrt{3} <
\frac{4}{3}$, allowing the models with $q=4$ and $q=2$ to have the same
dimension.
\begin{table*}
\caption{FK and geometrical fractal dimensions
characterizing the $q$-state Potts models, with $q=0,1,2,3,4$.
\label{table:Ds}}
\begin{tabular}{c|cccccccccccc}
\hline \hline & & & &  & & & & & &  \\[-.7cm] 
 $q$ & $c$ & $m$ & $\bar\kappa$  & $D_{\rm C}$ & $D_{\rm H}$ & $D_{\rm EP}$ &
 $y_T$ & $D_{\rm RB}$ & $D_{\rm C}^{\rm G}$ & $D_{\rm H}^{\rm G}$ & $y^{\rm
 G}_T$ & $D_{\rm RB}^{\rm G}$  \\[.1cm]
\hline & & & & & & & & & &   \\[-.7cm] 
 0 & $-2$ & 1 & $2$  & $2$ & $2$ & $\frac{5}{4}$ & $0$ & $\frac{5}{4}$ &
 $\frac{35}{16}$ & $\frac{5}{4}$ & $\frac{9}{4}$ & $-\frac{7}{4}$ \\
 1 & $0$ & 2 & $\frac{3}{2}$ & $\frac{91}{48}$ & $\frac{7}{4}$ &
 $\frac{4}{3}$ & 
 $\frac{3}{4}$ & $\frac{3}{4}$ & $2$ & $\frac{4}{3}$ & $2$  &
 $-\frac{11}{12}$  \\ 
 2 & $\frac{1}{2}$ & 3 & $\frac{4}{3}$ & $\frac{15}{8}$ & $\frac{5}{3}$
 & $\frac{11}{8}$ & $1$ & $\frac{13}{24}$ & $\frac{187}{96}$ &
 $\frac{11}{8}$ & $\frac{15}{8}$ & $-\frac{5}{8}$ \\
 3 & $\frac{4}{5}$ & 5 & $\frac{6}{5}$ & $\frac{28}{15}$ &
 $\frac{8}{5}$ & $\frac{17}{12}$ & $\frac{6}{5}$ & $\frac{7}{20}$ &
 $\frac{153}{80}$ & $\frac{17}{12}$ & $\frac{7}{4}$ & $-\frac{23}{60}$ \\
 4 & $1$ & $\infty$ & 1  & $\frac{15}{8}$ & $\frac{3}{2}$ &
 $\frac{3}{2}$ & $\frac{3}{2}$ & $0$ & $\frac{15}{8}$ & $\frac{3}{2}$ &
 $\frac{3}{2}$ & $0$  \\[.1cm] \hline \hline
\end{tabular}
\end{table*}

The physical meaning of the thermal eigenvalue $y^{\rm G}_T$ is related
to the existence of a critical magnetic field $H_\mathrm{s}(K)$, such
that in the region $K<K_\mathrm{c}$, $H > H_\mathrm{s}(K)$ a geometrical
cluster spanning the lattice (hence the subscript ``s'') is always
present.  The eigenvalue $y^{\rm G}_T$ namely determines the vanishing
of this field as $K$ approaches $K_\mathrm{c}$ from below
\cite{StellaVdzandePRL}:
\begin{equation} 
H_\mathrm{s}(K) \sim (K_\mathrm{c} - K)^{y^{\rm G}_T}.
\end{equation}

\section{Tricritical Potts Model}
\label{sec:tri}
In the traditional representation in terms of the parameter $m$
determining the central charge $c$ through Eq.~(\ref{cm}), the $q$-state
tricritical Potts models is parameterized by \cite{Cardyrev}
\begin{equation} 
\sqrt{q} = 2 \cos \left( \frac{\pi}{m} \right) =  - 2 \cos
\left( \frac{1+m}{m} \pi \right) ,
\end{equation} 
where $m$ is restricted to $1 \leq m \leq \infty$. As for the Potts
models, usually only the first equation is given.  The second one is
included because on comparing with the parameterization (\ref{qmPotts})
of the Potts models, we observe that the two are related by inverting
$m/(1+m)$.  Given the connection (\ref{kappam}) with the $\bar\kappa$
notation, it follows that this is nothing but the central-charge
conserving map $\bar\kappa \to 1/\bar\kappa$ which relates the FK and
geometrical clusters of a given Potts model.  Whereas $\bar\kappa_+$
parameterizes the Potts branch (\ref{Potts_branch}), the solution
$\bar\kappa_-$ of Eq.~(\ref{kappac}) parameterizes the tricritical
branch,
\begin{equation} 
\label{partri}
\sqrt{q} = -2 \cos (\pi/\bar\kappa_-),
\end{equation} 
with $\bar\kappa_-$ restricted to the values $\frac{1}{2} \le
\bar\kappa_- \le 1$, so that the argument of the cosine now takes values
in the interval $[\pi,3\pi]$.  (From now on, the $\bar\kappa$'s are
given a subscript plus or minus to indicate the solution larger or
smaller than 1.  Up to this point only the larger solution was used and
no index was needed to distinguish.)  The results obtained for the Potts
models can be simply transcribed to the tricritical models, provided
$\bar\kappa_+$ used on the Potts branch is replaced by $\bar\kappa_-$.
This close relation between the two models was first observed by
Nienhuis {\it et al.} \cite{NBRS}.  Note that for $\frac{1}{2} \leq
\bar\kappa_- <\frac{2}{3}$ the geometrical cluster dimension exceeds the
available number of dimensions, which is unphysical.  The eigenvalue
$y^{\rm G}_T$ (\ref{yGT}) with $\bar\kappa=\bar\kappa_+=1/\bar\kappa_-$
is one of the two leading thermal eigenvalues of the tricritical Potts
model \cite{Nienhuis}.  The second one is given by the inverse
correlation length exponent $1/\nu$, with
\begin{equation} 
\label{nutri}
\nu = \frac{1}{4} \frac{1}{1-\bar\kappa_-}.
\end{equation} 

Because replacing $\bar\kappa_+$ with $\bar\kappa_-$ is tantamount to
replacing $\bar\kappa_+$ with $1/\bar\kappa_+$, it follows that for
given central charge, the FK clusters of the tricritical model are the
geometrical clusters of the Potts model (and {\it vice versa}).  For
example, the fractal dimension $D_\mathrm{C}$ of the FK clusters in the
critical regime, $D_\mathrm{C} = 1 + 1/2\bar\kappa_+ + 3
\bar\kappa_+/8$, translates into $1 + 1/2\bar\kappa_- + 3 \bar\kappa_-/8
= 1 + \bar\kappa_+/2 + 3/8 \bar\kappa_+$ for the tricritical regime.
This is precisely the fractal dimension (\ref{DgeoD}) of the geometrical
Potts clusters.

As $q$ increases, the critical and tricritical points approach each
other until they annihilate at the critical value $q=4$, where the FK
and geometrical clusters coincide and the red bond dimension vanishes.
As stressed by Coniglio \cite{Coniglio1989}, the vanishing of $D_{\rm
RB}$ signals a drastic change in the fractal structure, anticipating a
first-order phase transition.

\section{High-Temperature Representation}
\label{sec:HT}
\subsection{Monte Carlo Study}
\begin{figure}
\begin{center}
\psfrag{b}[t][t][1][0]{}
\psfrag{=}[t][t][1][0]{}
\psfrag{0.95}[t][t][1][0]{}
\psfrag{4.00}[t][t][1][0]{}
\includegraphics[width=3.5cm, height=4.0cm]{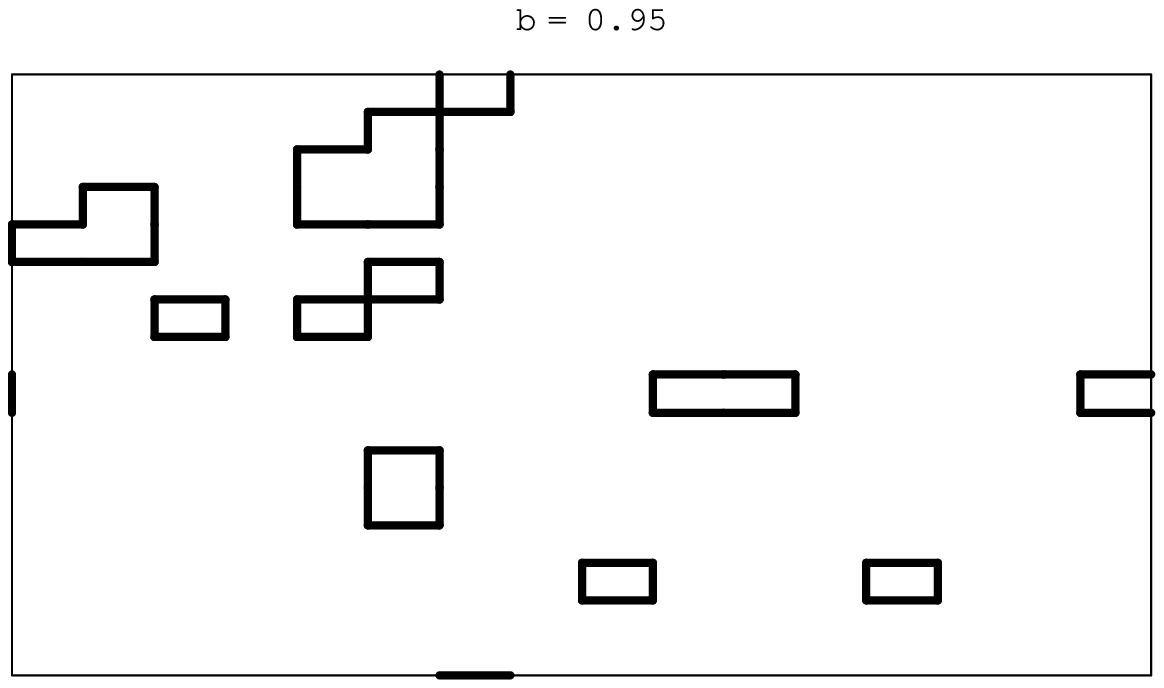} \hspace{1.5cm}
\includegraphics[width=3.5cm, height=4.0cm]{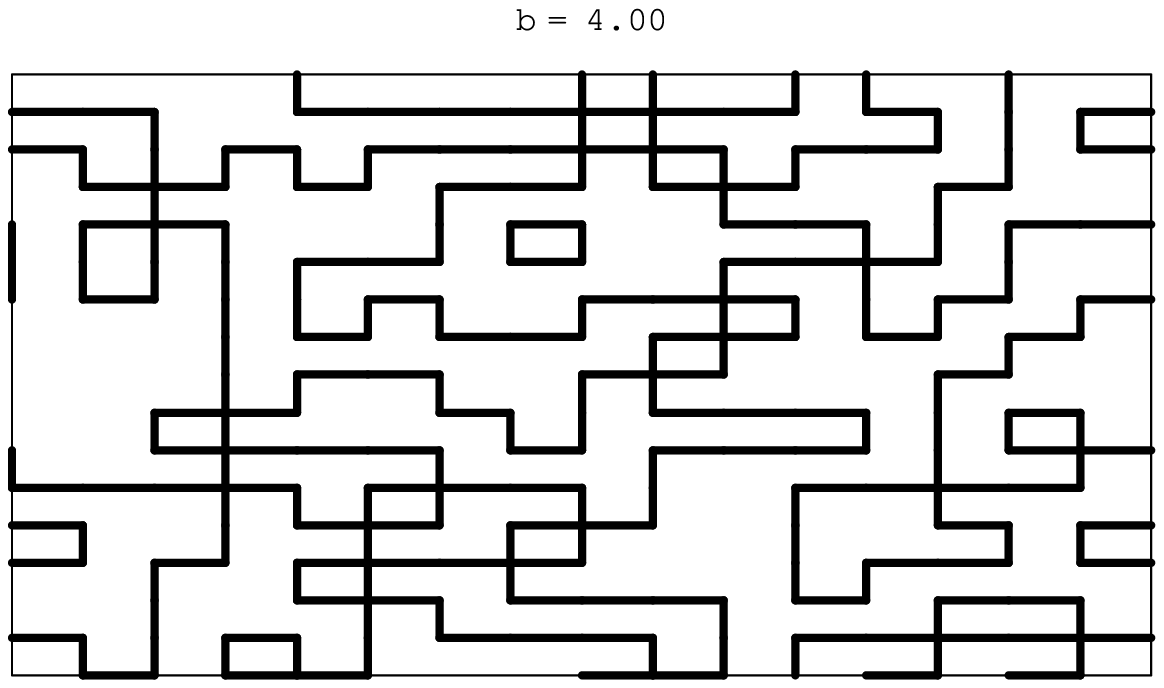} 
\end{center}
\caption{Typical graph configurations, somewhat resembling the oil
  painting {\it Rhythm of a Russian Dance} by {\sf De Stijl} artist Theo
  van Doesburg (1883-1931), generated on a $16 \times 16$ square lattice
  with periodic boundary conditions in the high- (left panel) and
  low-temperature (right panel) phase.
  \label{fig:vis}}
\end{figure}

\begin{figure}
\begin{center}
\psfrag{s}[t][t][.8][0]{selected plaquette}
\includegraphics[width=4.5cm]{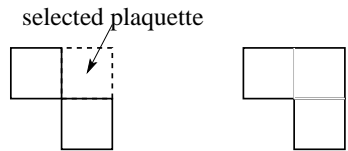}
\end{center}
\caption{Update mechanism at work. {\it Left panel}: Present graph
  with the plaquette selected for updating indicated by the broken
  square.  {\it Right panel}: New graph after the update proposal is
  accepted.  Both the old and new graph consist of 8 bonds, in
  accordance with Eq.~(\ref{ll}) since two bonds on the plaquette were
  already occupied.
  \label{fig:update}}
\end{figure}

To support the picture discussed above we carry out a Monte Carlo
simulation of the high-temperature (HT) representation of the 2-state
Potts, i.e., Ising model, adopting a new update algorith
\cite{Erkinger}.  HT, or strong coupling expansions can be visualized by
graphs on the lattice, with each occupied bond representing a certain
contribution to the partition function.  For the Ising model, defined by
the Hamiltonian
\begin{equation} 
-\beta \mathcal{H} = \beta \sum_{\langle ij\rangle} S_i S_j, \quad S_i = \pm 1,
\end{equation} 
where the coupling constant is taken to be unity, the HT representation
reads \cite{Feynman}:
\begin{equation} 
\label{HTIsing}
Z = (\cosh \beta)^{2N} 2^N \sum_{\{\Gamma_{\rm O}\}} v^b,
\end{equation}
where $\{\Gamma_{\rm O}\}$ denotes the set of \textit{closed} graphs
specified by $b$ occupied bonds, and $v = \tanh \beta$.  Traditionally,
HT expansions are carried out exactly up to a given order by enumerating
all possible ways graphs up to that order can be drawn on the lattice.
We instead generate possible graph configurations by means of a
Metropolis update algorithm, involving single plaquettes \cite{Erkinger}
(see Fig.~\ref{fig:vis} for typical configurations generated in the low-
and high-temperature phases).  By taking plaquettes as building blocks,
the resulting HT graphs are automatically closed--as required
\cite{Feynman}.  The update is such that all the bonds of a selected
plaquette are changed, i.e., those that were occupied become unoccupied
and \textit{vice versa} \cite{Erkinger} (see Fig.~\ref{fig:update}).  A
proposed update resulting in $b'$ occupied bonds is accepted with
probability
\begin{equation} 
p_\mathrm{HT} = \min \left(1,v^{b'-b} \right),
\end{equation} 
where $b$ denotes the number of occupied bonds before the update.  With
$l$ denoting the number of bonds on the plaquette already occupied, $b$
and $b'$ are related through
\begin{equation} 
\label{ll}
b' = b + 4 - 2 l.
\end{equation}  

In the following, we focus exclusively on the graphs and measure typical
cluster quantities, such as the graph distribution, the size of the
largest graph, and whether or not a graph spans the lattice.  From this,
the temperature where the graphs proliferate as well as their fractal
dimension can be extracted as in percolation theory.  Both the
proliferation temperature and the associated correlation length exponent
turn out to coincide with their Ising counterparts.

By the well-known Kramers-Wannier duality \cite{KrWa}, the HT graphs
form Peierls domain walls \cite{Peierls} separating spin clusters of
opposite orientation on the dual lattice.  Each bond in a HT graph
intersects a nearest neighbor pair on the dual lattice of unlike spins
perpendicular to it.  In other words, the HT graphs are the boundaries
of {\it geometrical} spin clusters (albeit on the dual lattice) whose
fractal dimension we wish to establish.  The advantage of the plaquette
update we use is that these boundaries are simulated directly without
first considering the corresponding cluster.  At the critical
temperature, the domain walls lose their line tension and proliferate.

When interpreted as domain walls, the HT graphs should strictly speaking
be cut at the vertices, so that the graphs break down in separate
polygons without self-intersections that only touch at the corners where
the vertices were located.  However, it is expected that this does not
change the universal properties at criticality we wish to determine.

\begin{figure}
\begin{center}
\includegraphics[width=2.cm]{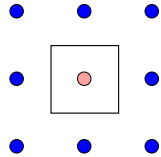}
\end{center}
\caption{A plaquette on the original lattice corresponds to a Peierls domain
  wall on the dual lattice, separating the site at its center with
  reversed spin from the outside.
  \label{fig:plaquette}}
\end{figure}

From the duality argument it also follows that the plaquette update is
equivalent to a single spin update on the dual lattice (see
Fig.~\ref{fig:plaquette}).
To illustrate this, the internal energy $U$ is computed, using the
plaquette update.  On an infinite lattice, the Kramers-Wannier duality
implies that observables calculated at an inverse temperature $\beta$ in
the original Ising model can be transcribed to those of the dual model
at an inverse temperature $\tilde\beta$.  The relation between the two
temperatures follows from noting that, an occupied HT bond represents a
factor $v$, while a nearest neighbor pair on the dual lattice of
unlike spin on each side of the HT bond carries a Boltzmann weight
$\exp(-2\tilde \beta)$, so that \cite{KrWa}
\begin{equation} 
\label{bb}
\tanh \beta = {\rm e}^{-2\tilde \beta},
\end{equation} 
or $\sinh 2 \beta = 1/\sinh 2 \tilde\beta$.  

\begin{figure}
\begin{center}
\psfrag{x}[t][t][.8][0]{$\beta$}
\psfrag{y}[t][t][.8][0]{$U$}
\psfrag{Ising}[c][t][.8][0]{Ising}
\psfrag{MC}[c][t][.8][0]{MC}
\psfrag{Dual}[c][t][.8][0]{Dual}
\includegraphics[width=10.cm]{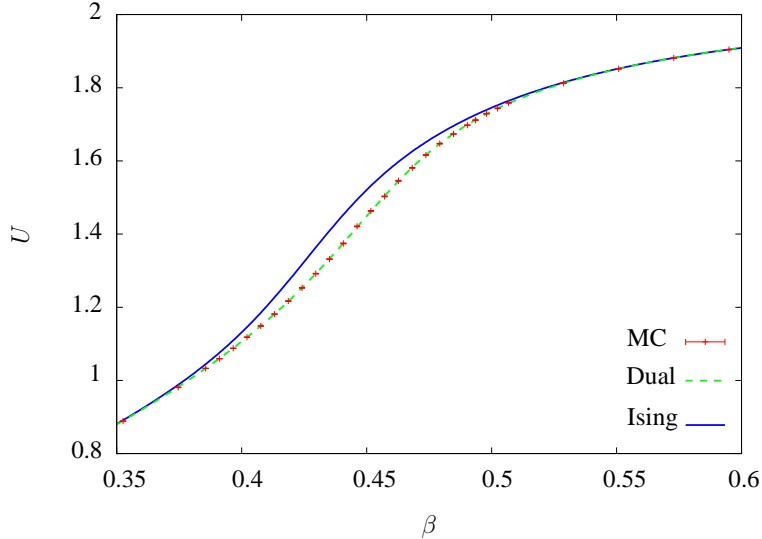}
\end{center}
\caption{Exact internal energy $U$ on a $16^2$ lattice for the Ising and
dual model [transcribed to the original model using Eq.~(\ref{bb})], and
the Monte Carlo data obtained using the plaquette update.
  \label{fig:dual_update}}
\end{figure}

On a finite lattice with periodic boundary conditions, however, a
mismatch arises because {\it single} HT graphs wrapping the (finite)
lattice are not generated by the plaquette update--such graphs always
come in pairs.  The HT Monte Carlo study will therefore not exactly
simulate the Ising model with periodic boundary conditions, at least not
for small lattice sizes.  (For larger lattices, single graphs wrapping
the lattice become highly unlikely, so that their absence will not be
noticed anymore, and the HT Monte Carlo simulation becomes increasingly
more accurate.)  In contrast, on the dual side, where the plaquette
update corresponds to a single spin update, this class of graphs is not
compatible with the periodic boundary conditions, so that they should
not be included.  Hence, the plaquette update simulates the
(transcribed) dual rather than the original model itself.
Figure~\ref{fig:dual_update} gives the exact internal energy of the
original Ising model on a finite lattice with periodic boundary
conditions \cite{FeFi,CompPhys} and that of the dual model transcribed
to the original one using Eq.~(\ref{bb}).  In the figure, also the data
points obtained using the plaquette update are included and seen to
indeed coincide with the dual curve.  For increasing lattice sizes, the
dual and the original curves approach each other.
\subsection{Simulation}

\begin{figure}
\begin{center}
\psfrag{x}[t][t][.8][0]{$\beta/\beta_{\rm c}$}
\psfrag{y}[t][t][.8][0]{$P_{\rm S}$}
\psfrag{Ps16}[t][t][.8][0]{$16$}
\psfrag{Ps32}[t][t][.8][0]{$32$}
\psfrag{Ps64}[t][t][.8][0]{$64$}
\psfrag{Ps128}[t][t][.8][0]{$128$}
\psfrag{Ps256}[t][t][.8][0]{$256$}
\includegraphics[width=10.cm]{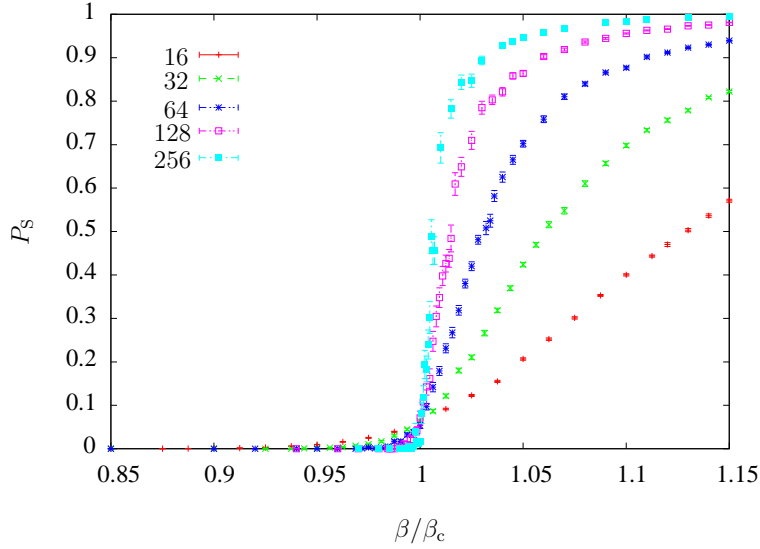}
\end{center}
\caption{Probability $P_{\rm S}$ for the presence of a spanning graph as
function of $\beta$ measured for different lattice sizes $L$.  Within the
achieved accuracy, the curves cross at the thermal critical point $\beta
= \beta_{\rm c}$.
  \label{fig:cross}}
\end{figure}

To determine the graph proliferation temperature, the probability
$P_{\rm S}$ for the presence of a graph spanning the lattice as function
of $\beta$ is measured for different lattice sizes
\cite{StauferAharony}.  For small $\beta$, $P_{\rm S}$ tends to zero,
while for large $\beta$ it tends to unity.  We consider a graph spanning
the lattice already when it does so in just one direction.  Ideally, the
curves obtained for different lattice sizes cross in a single point,
marking the proliferation temperature.  It is seen from
Fig.~\ref{fig:cross} that within the achieved accuracy, the measured
curves cross at the thermal critical point, implying that the HT graphs
(domain walls) lose their line tension and proliferate precisely at the
Curie point.

The data was collected in $3.3 \times 10^5$ Monte Carlo sweeps of the
lattice close to the critical point and $1.1 \times 10^5$ outside the
critical region, with about 10\% of the sweeps used for equilibration.
After each sweep, the resulting graph configuration was analyzed.
Statistical errors were estimated by means of binning.

\begin{figure}
\begin{center}
\psfrag{x}[t][t][.8][0]{$(\beta/\beta_{\rm c}-1)L^{1/\nu}$}
\psfrag{y}[t][t][.8][0]{$P_{\rm S}$}
\psfrag{Ps16}[t][t][.8][0]{$16$}
\psfrag{Ps32}[t][t][.8][0]{$32$}
\psfrag{Ps64}[t][t][.8][0]{$64$}
\psfrag{Ps128}[t][t][.8][0]{$128$}
\psfrag{Ps256}[t][t][.8][0]{$256$}
\includegraphics[width=10.cm]{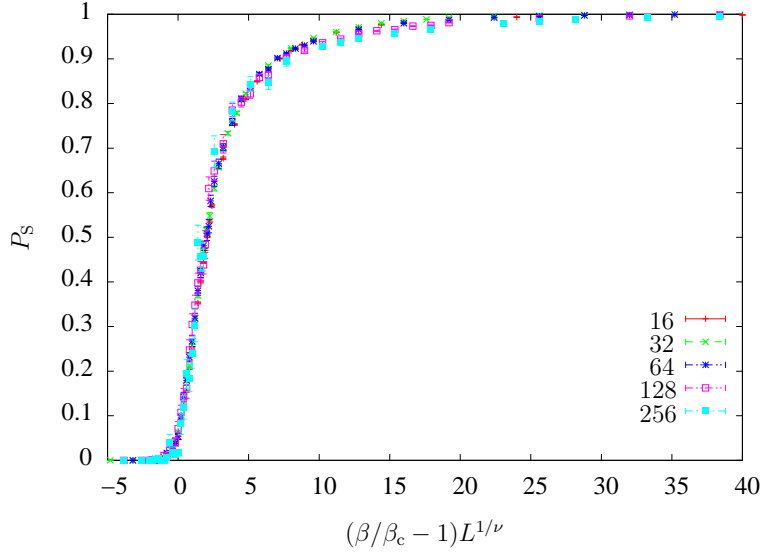}
\end{center}
\caption{The raw data of Fig.~\ref{fig:cross} replotted as function of
$(\beta/\beta_{\rm c}-1)L^{1/\nu}$, with the Ising choice $\nu=1$.  The
data collapse is satisfactory over the entire temperature range.
  \label{fig:Ps_collapse}}
\end{figure}

Finite-size scaling \cite{StauferAharony} predicts that the raw $P_{\rm
  S}$ data obtained for different lattice sizes collapse onto a single
curve when plotted as function of $(\beta/\beta_{\rm c}-1)L^{1/\nu}$
with the right choice of the exponent $\nu$.  By duality, the relevant
correlation length here is that of the Ising model, so that $\nu$ takes
the Ising value $\nu=1$.  With this choice, a satisfying collapse of the
data is achieved over the entire temperature range (see
Fig.~\ref{fig:Ps_collapse}).

\begin{figure}
\begin{center}
\psfrag{x}[t][t][.8][0]{$\beta/\beta_\mathrm{c}$}
\psfrag{y}[t][t][.8][0]{$P_\infty$}
\psfrag{16}[t][t][.8][0]{$16$}
\psfrag{32}[t][t][.8][0]{$32$}
\psfrag{64}[t][t][.8][0]{$64$}
\psfrag{128}[t][t][.8][0]{$128$}
\psfrag{256}[t][t][.8][0]{$256$}
\includegraphics[width=10.cm]{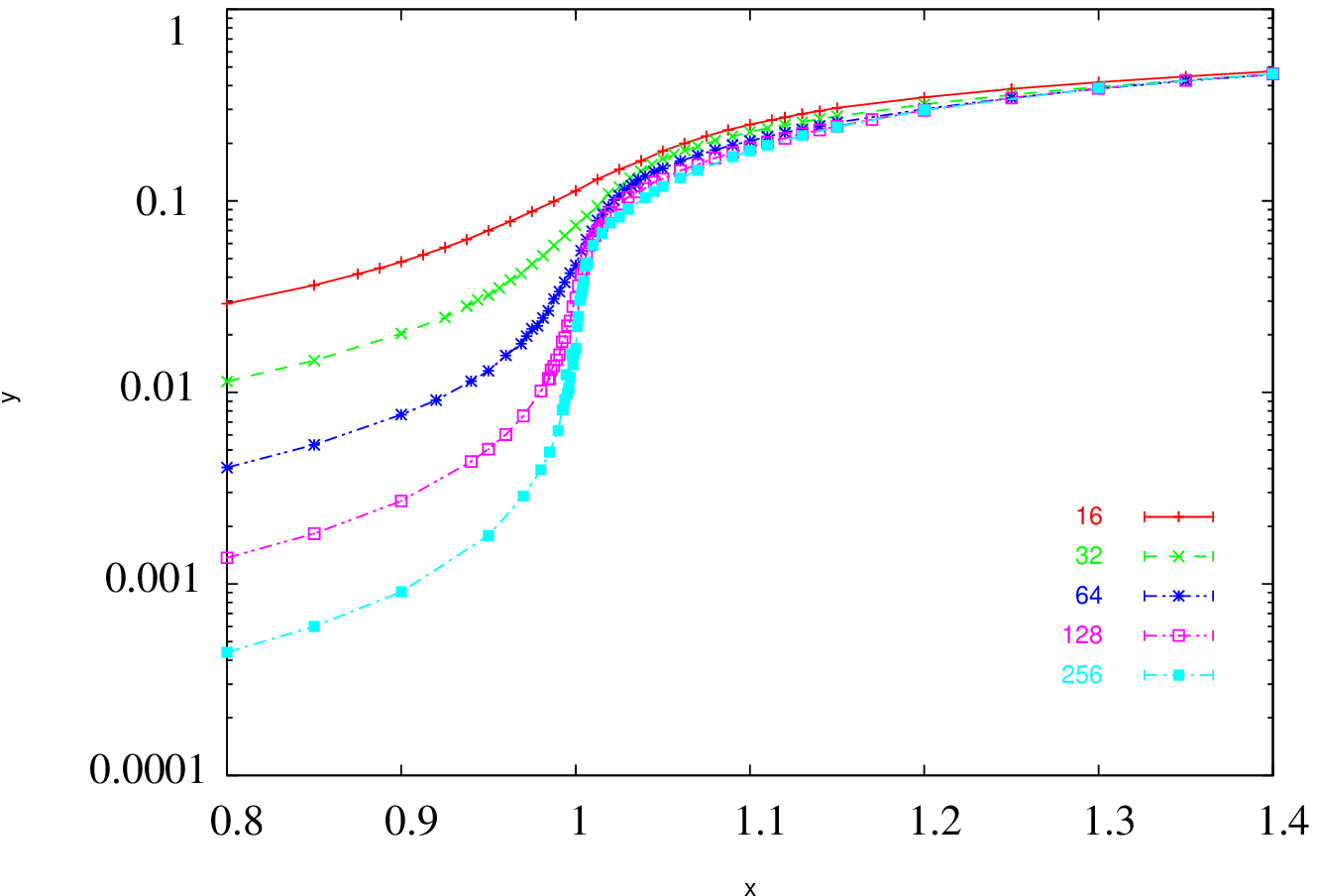} \\[1.cm]
\psfrag{x}[t][t][.8][0]{$\beta/\beta_\mathrm{c}$}
\psfrag{y}[t][t][.8][0]{$\chi_\mathrm{G}$}
\includegraphics[width=10.cm]{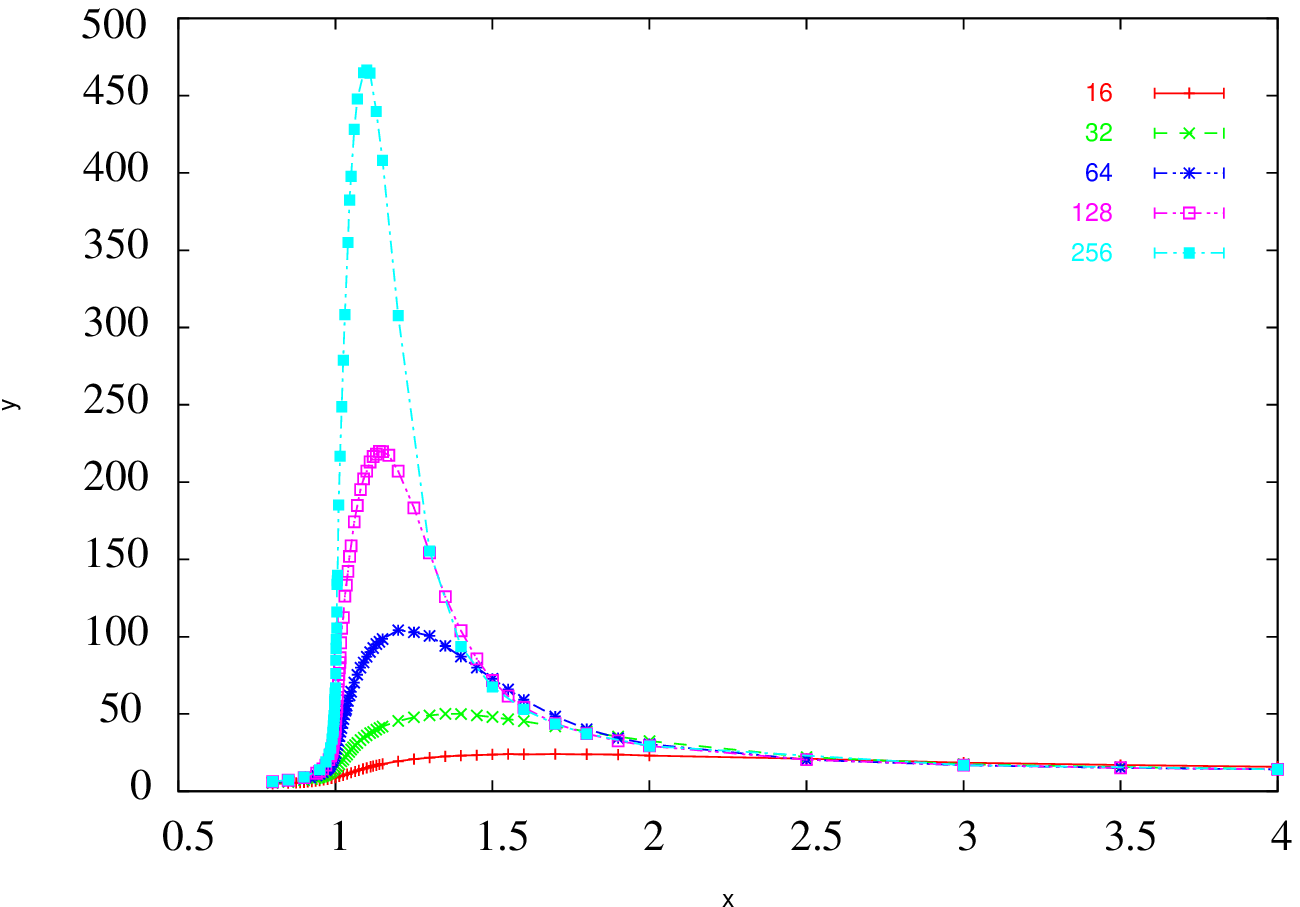}
\end{center}
\caption{The percolation strength $P_\infty$ (top panel) and average
  graph size $\chi_\mathrm{G}$ (bottom panel) as function of
  $\beta/\beta_\mathrm{c}$ for different lattice sizes $L$.
  \label{fig:Pochi}}
\end{figure}

\begin{figure}
\begin{center}
\psfrag{x}[t][t][.8][0]{$L$}
\psfrag{y}[t][t][.8][0]{$P_\infty$}
\psfrag{f}[t][t][.8][0]{} 
\includegraphics[width=10.cm]{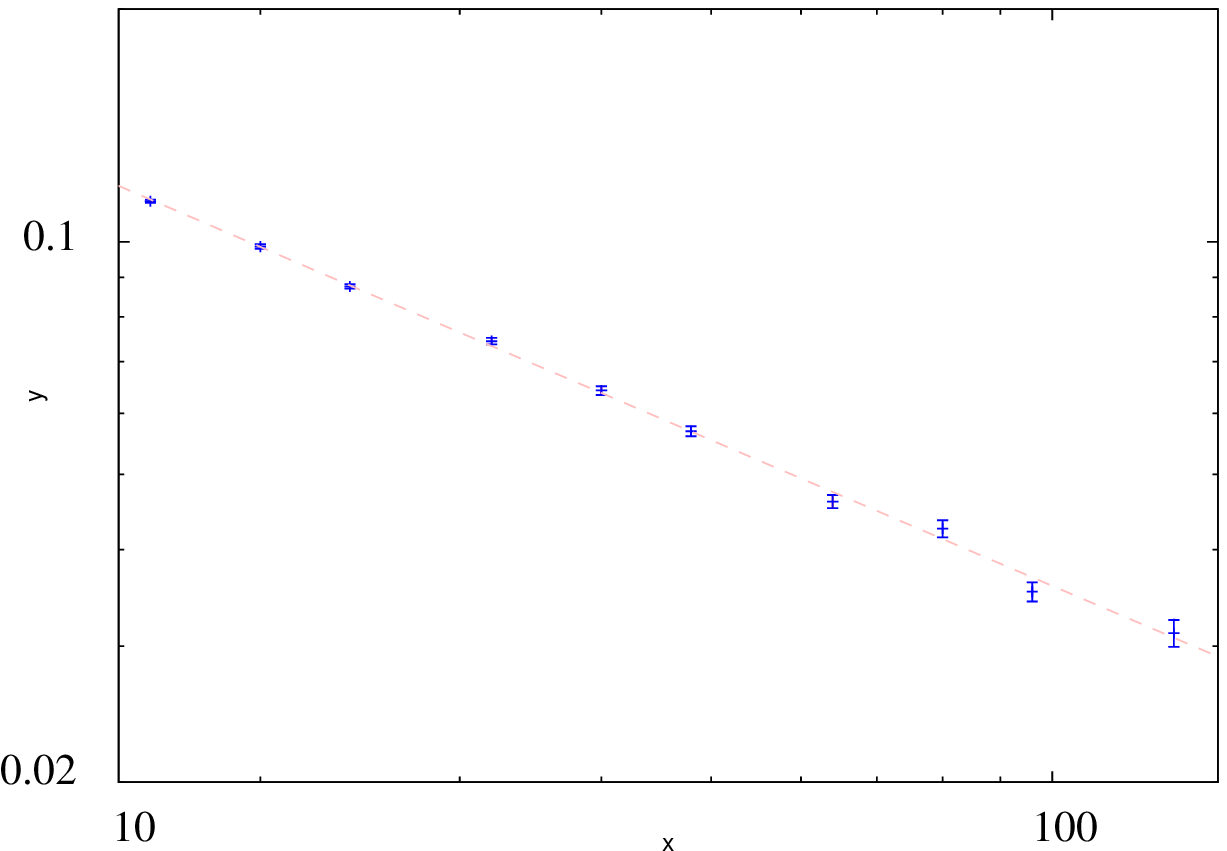} \\[1.cm]
\psfrag{x}[t][t][.8][0]{$L$}
\psfrag{y}[t][t][.8][0]{$\chi_\mathrm{G}$}
\psfrag{f}[t][t][.8][0]{} 
\includegraphics[width=10.cm]{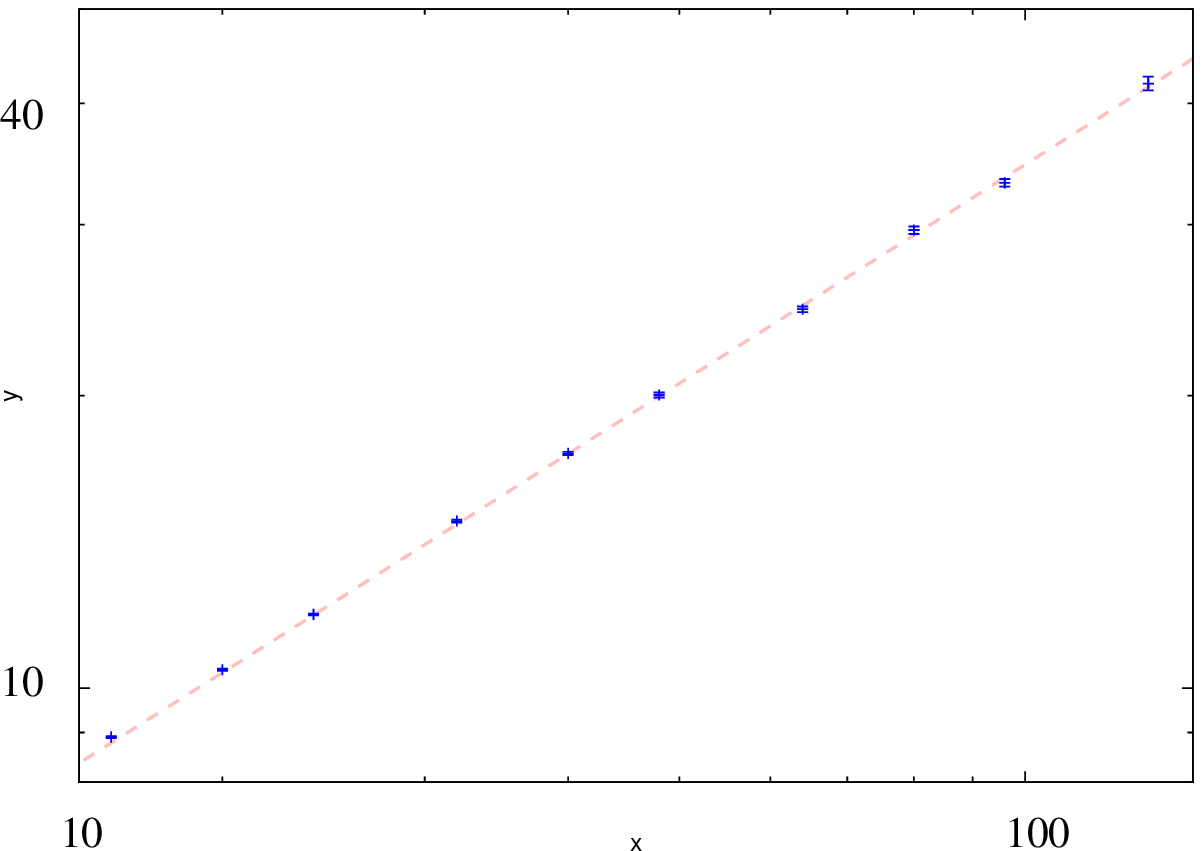}
\end{center}
\caption{Log-log plot of the percolation strength $P_\infty$ (top panel)
  and average graph size $\chi_\mathrm{G}$ (bottom panel) at the critical
  temperature as function of the lattice size $L$.  The straight lines
  $0.642 \, L^{-0.626}$ and $1.139 \, L^{0.740}$, respectively are
  obtained through two-parameter fits.
  \label{fig:fit}}
\end{figure}

Next, the cluster exponents $\sigma_{\rm G}$ and $\tau_{\rm G}$
specifying the graph distribution,
\begin{equation} 
\ell_b \sim b^{- \tau_{\rm G}} {\rm e}^{- \theta b}, \quad \theta \propto
(\beta-\beta_{\rm c})^{1/\sigma_{\rm G}}
\end{equation} 
are determined, where $\ell_b$ denotes the average number density of
graphs containing $b$ bonds.  To this end we measure the so-called
percolation strength $P_\infty$, giving the fraction of bonds in the
largest graph, and as second independent observable the average graph
size \cite{StauferAharony} (see Fig.~\ref{fig:Pochi})
\begin{equation} 
\chi_\mathrm{G} = \frac{\sum_b' b^2 \ell_b}{\sum_b' b \ell_b},
\end{equation} 
where the prime on the sum indicates that the largest graph in each
measurement is omitted.
Close to the proliferation temperature, these observables obey the
finite-size scaling relations \cite{BinderHeermann}
\begin{equation} 
P_\infty = L^{-\beta_{\rm G}/\nu} \, {\sf P}(L/\xi), \quad
\chi_\mathrm{G} = L^{\gamma_{\rm G}/\nu} \, {\sf X} (L/\xi),
\end{equation} 
where $\xi$ is the correlation length and the critical exponents
$\beta_{\rm G}, \gamma_{\rm G}$ are related to $\sigma_{\rm G},
\tau_{\rm G}$ through Eq.~(\ref{perce}) written in terms of the
variables appropriate for the graph exponents.  Precisely at $T_{\rm
  c}$, these scaling relations imply an algebraic dependence on the
system size $L$, allowing for a determination of the exponent ratios
(see Table \ref{table:num} and Fig.~\ref{fig:fit}).

\begin{table*}
\caption{Percolation strength $P_\infty$ and average graph size
$\chi_\mathrm{C}$ at the (inverse) critical temperature $\beta_{\rm c} =
\ln(1+\sqrt{2})/2 = 0.440687\cdots$ for various lattice sizes $L$ (see
Fig.\ \ref{fig:fit}). \label{table:num}}
{\tiny
\begin{tabular}{c|cccccccccc}
\hline \hline & & & &  & & & & & &  \\[-.3cm] 
$L$ & 16 & 20 & 24 & 32 & 40 & 48 & 64 & 80 & 96 & 128 \\[.2cm]
\hline & & & & & & & & & &   \\[-.3cm] 
$P_\infty$ & 0.1129(6) & 0.0986(7) & 0.0876(6) & 0.0744(7) & 0.0642(8) & 0.0569(8) &
0.0461(9) & 0.0426(11) & 0.0353(10) & 0.0312(12) \\
$\chi_\mathrm{C}$ & 8.90(4) & 10.44(4) & 11.91(5) & 14.86(8) & 17.44(10) &
20.03(16) & 24.55(22) & 29.61(34) & 33.13(39) & 41.93(81) \\[.1cm] \hline \hline
\end{tabular}
}
\end{table*}

The data was fitted using the nonlinear least-squares
Marquardt-Levenberg algorithm, giving
\begin{equation} 
\label{res}
\beta_{\rm G} = 0.626(7) \quad \gamma_{\rm G} = 0.740(4) ,
\end{equation} 
with $\chi^2/{\rm d.o.f} = 1.15$ and $0.94$, respectively, and where it
was used that $\nu =1$.  These values are perfectly consistent with the
fractions $\beta_{\rm G}=\frac{5}{8}=0.625, \gamma_{\rm G}
=\frac{3}{4}=0.75$, leading to the exponents
\begin{equation} 
\sigma_{\rm G} = \frac{8}{11}, \quad \tau_{\rm G} = \frac{27}{11},
\end{equation} 
and the fractal dimension
\begin{equation} 
\label{DHG}
D_{\rm H}^{\rm G} = \frac{11}{8}
\end{equation} 
of the HT graphs we were seeking.  By duality, this fractal dimension
equals that of the hull bounding the geometrical spin clusters (Peierls
domain walls) on the dual lattice.  Our numerical result agrees
with Eq.~(\ref{DGH}) with $\bar\kappa=\bar\kappa_{\rm Ising}=
\frac{4}{3}$ appropriate for the Ising model.  

The value (\ref{DHG}) was predicted by Duplantier and Saleur, using the
Coulomb gas map \cite{DS88}.  Subsequent support for that prediction was
provided by Vanderzande and Stella \cite{VdzandeStellaJP} who drew on
earlier numerical work by Cambier and Nauenberg \cite{CaNau}.  A first
direct numerical determination was given by Dotsenko {\it et al.}
\cite{Dotsenkoetal}.  Employing the Swendsen-Wang cluster update, these
authors analyzed the geometrical spin clusters and their hulls at the
critical temperature.  They extracted the fractal dimension from the
resulting hull distribution, which is algebraic at the critical
temperature.  Apart from directly simulating the hulls with the
plaquette update, another advantage of our approach is the use of
finite-size scaling which is generally considered more reliable than the
extraction of exponents by fits to the data obtained for a fixed lattice
size.

As argued in Sec.~\ref{sec:tri} for the general case, the geometrical
Ising clusters correspond to the tricritical Potts model with
$\bar\kappa = \bar\kappa_- = 1/\bar\kappa_{\rm Ising}=\frac{3}{4}$,
which according to Eq.~(\ref{partri}) is the tricritical $q=1$ model
\cite{StellaVdzandePRL}.  In addition to the correlation length exponent
$\nu=1$ which we, in accordance with Eq.~(\ref{nutri}), observed
numerically and the second thermal eigenvalue
$y_T^\mathrm{G}=\frac{15}{8}$, this tricritical behavior is further
characterized by \cite{Nienhuis} $\beta_\mathrm{C} = \frac{5}{96}$.
This value follows from the scaling relation (\ref{yH}) with the fractal
dimension $D_\mathrm{C}$ of the FK clusters replaced by that of their
geometrical counterpart (\ref{DgeoD}) with $\bar\kappa=\bar\kappa_{\rm
  Ising}= \frac{4}{3}$ \cite{StellaVdzandePRL}.

The fact that the two correlation length exponents featuring in the
critical $q=2$ and tricritical $q=1$ Potts models are equal is special
to this case, being a result of the Kramers-Wannier duality.  Indeed,
equating the correlation length exponent (\ref{cesPotts}) of the
critical Potts models and that of the tricritical Potts models given in
(\ref{nutri}) with $\bar\kappa_- = 1/\bar\kappa$ to assure that both
models have the same central charge, yields $\bar\kappa=\bar\kappa_{\rm
Ising}= \frac{4}{3}$ as only physical solution.

\section{Summary}
\label{sec:summary}
In this paper, it is shown that the geometrical spin clusters of the
pure $q$-state Potts model in two dimensions encode the tricritical
behavior of the site diluted model.  These clusters, formed by nearest
neighbor sites of like spins, were shown to be mirror images of FK
clusters, which in turn encode the critical behavior.  Since the mirror
map conserves the central charge, both cluster types (and thus both
fixed points) are in the same universality class.  The geometrical
picture was supported by a Monte Carlo simulation of the
high-temperature representation of the Ising model, corresponding to
$q=2$.  The use of a plaquette update allowed us to directly simulate
the hulls of the geometrical clusters and to accurately determine their
fractal dimension.

\ack The authors acknowledge helpful discussions and correspondence with
H. Kleinert and C. Vanderzande.  This work is partially supported by the
EC research network HPRN-CT-1999-00161 EUROGRID--"Discrete Random
Geometries: from solid state physics to quantum gravity" and by the
German-Israel Foundation (GIF) under grant No.\ I-653-181.14/1999.  AS
gratefully acknowledges support by the DFG through the Graduiertenkolleg
``Quantenfeldtheorie'' and the Theoretical Sciences Center (NTZ) of the
Universit\"at Leipzig.  The project is carried out on computers running
GNU/Linux.


\begin{thebibliography}{99} 
\bibitem{Potts} R. B. Potts, Proc. Camb. Phil. Soc. {\bf 48}, 106 (1952).
\bibitem{FK} C. M. Fortuin and P. W.  Kasteleyn, Physica {\bf 57}, 536
  (1972).
\bibitem{Wu} F. Y. Wu, Rev. Mod. Phys. {\bf 54}, 235 (1982).
\bibitem{StauferAharony} D. Stauffer and A. Aharony, {\it Introduction
to Percolation Theory}, 2nd edition (Taylor \& Francis, London, 1994).
\bibitem{SwendsenWang} R. H. Swendsen and J. S. Wang,
Phys. Rev. Lett. {\bf 58}, 86 (1987).
\bibitem{Wolff} U. Wolff, Phys. Rev. Lett. {\bf 62}, 361 (1989).
\bibitem{Baxter} R. J. Baxter, J. Phys. C {\bf 6}, L445 (1973).
\bibitem{NBRS} B. Nienhuis, A. N. Berker, E. K. Riedel, and M. Schick,
  Phys. Rev. Lett. {\bf 43}, 737 (1979).
\bibitem{Schramm} O. Schramm, {\em Israel J. Math.}, {\bf 118}, 221
(2000).
\bibitem{footnote} The parameter $\bar\kappa$ we use is related to the
label $\kappa$ in Ref.~\cite{Schramm} through $\bar\kappa=\kappa/4$.
The reason for our convention will become clear when we proceed [see,
e.g., Eq.~(\ref{constraint})].
\bibitem{denNijs} M. P. M. den Nijs, J. Phys. A {\bf 12}, 1857 (1979);
Phys. Rev. B {\bf 27}, 1674 (1983).
\bibitem{Nienhuis} B. Nienhuis, J. Phys. A {\bf 15}, 199 (1982); in:
{\it Phase Transitions and Critical Phenomena}, edited by C. Domb and
J. L. Lebowitz (Academic, London, 1987), Vol. 11, p.1.
\bibitem{SD}  H. Saleur and B. Duplantier, Phys. Rev. Lett. {\bf 58}, 2325
(1987).
\bibitem{Cardyrev} J. Cardy, in: {\it Phase Transitions and Critical
Phenomena}, edited by C. Domb and J. L. Lebowitz (Academic, London,
1987), Vol. 11, p.55
\bibitem{Duplantier02} B. Duplantier, J. Stat. Phys.  {\bf 110}, 691
  (2003).
\bibitem{RS} S. Rohde and O. Schramm, {\it Basic properties of SLE},
  Ann. Math. (to appear), {\tt math.PR/0106036} (2001).
\bibitem{BB} M. Bauer and D. Bernard, Commun. Math. Phys. {\bf 239}, 493
  (2003).
\bibitem{Duplantiermath} B. Duplantier, \emph{Conformal Fractal Geometry
    and Boundary Quantum Gravity}, \texttt{math-ph/0303034} (2003).
\bibitem{StellaVdzandePRL} A. L. Stella and C. Vanderzande,
  Phys. Rev. Lett. {\bf 62}, 1067 (1989).
\bibitem{VdzandeStellaJP} C. Vanderzande and A. L. Stella, J. Phys. A
{\bf 22}, L445 (1989).  

\bibitem{TeHe} T. Temesv\'ari and L. Her\'enyi, J. Phys. A {\bf 17},
1703 (1984).
\bibitem{ConKl} A. Coniglio and W. Klein, J. Phys. A {\bf 13}, 2775
  (1980).
\bibitem{Duplantier87} B. Duplantier, J. Stat. Phys. {\bf 49}, 411
(1987).
\bibitem{Erkinger} H.-M. Erkinger, {\it A new cluster algorithm for the
    Ising model}, {\it Diplomarbeit}, Technische Universit\"at Graz
  (2000).
\bibitem{DS88} B. Duplantier and H. Saleur, Phys. Rev. Lett. {\bf 61},
  1521 (1988).
\bibitem{Murata} K. K. Murata, J. Phys. A {\bf 12}, 81 (1979).
\bibitem{Vanderzande} C. Vanderzande, J. Phys. A {\bf 25}, L75 (1992).
  A typo seems to be present in the right hand of Eq.~(10) of that
  paper, giving the fractal dimension of the red bonds of the
  geometrical clusters.  It should read $(-4m-3)/2m(m+1)$, where $m$ is
  related to $\bar\kappa$ through Eq.~(\ref{kappam}).  The corresponding
  entries in Table~1 of that paper should be updated accordingly.
\bibitem{vortexland} A. M. J. Schakel, Phys.  Rev. E {\bf 63}, 026115
  (2001); J. Low Temp. Phys \textbf{129}, 323 (2002).
\bibitem{Coniglio1989} A. Coniglio, Phys. Rev. Lett. {\bf 62}, 3054
  (1989).
\bibitem{Duplantier00} B. Duplantier, Phys. Rev. Lett. {\bf 84}, 1363
(2000).
\bibitem{GrossmanAharony} T. Grossman and A. Aharony J. Phys. A {\bf
    19}, L745 (1986); {\bf 20}, L1193 (1987).
\bibitem{ABA} M. Aizenman, B. Duplantier, and A. Aharony, Phys. Rev.
  Lett. {\bf 83}, 1359 (1999).
\bibitem{AAMRH} J. Asikainen, A. Aharony, B. B. Mandelbrot, E. M. Rauch,
 and J.-P. Hovi, Eur. Phys. J. B \textbf{34}, 479 (2003).
\bibitem{Stanley} H. E. Stanley, J. Phys. A {\bf 10}, L211 (1977).
\bibitem{DS89} B. Duplantier and H. Saleur, Phys. Rev. Lett. {\bf 63},
  2536 (1989).
\bibitem{Feynman} R. P. Feynman, {\it Statistical Mechanics} (Benjamin,
  Reading, 1972).
\bibitem{KrWa} H. A. Kramers and G. H. Wannier, Phys. Rev. {\bf 60}, 252 (1941).
\bibitem{Peierls} R. Peierls, Proc. Camb. Phil. Soc. {\bf 32}, 477
(1936).
\bibitem{FeFi} B. Kaufman, Phys. Rev. \textbf{76}, 1232 (1949); A. E. Ferdinand
  and M. E. Fisher, Phys. Rev. {\bf 185}, 832 (1969).
\bibitem{CompPhys} W. Janke,
{\em Monte Carlo Simulations of Spin Systems\/},
in: {\em Computational Physics: Selected Methods -- Simple Exercises --
  Serious Applications\/}, edited by K. H. Hoffmann and M. Schreiber
(Springer, Berlin, 1996); p.~10.
\bibitem{BinderHeermann} K. Binder and D. W. Heermann, \textit{Monte
Carlo Simulation in Statistical Physics} (Springer, Berlin, 1997).
\bibitem{CaNau} J. L. Cambier and M. Nauenberg, Phys. Rev. B {\bf 34},
8071 (1986).
\bibitem{Dotsenkoetal} V. S. Dotsenko, M. Picco, P. Windey, G. Harris,
E. Martinec, and E. Marinari, Nucl. Phys. B {\bf 448} [FS], 577 (1995).



\end{thebibliography}
\end{document}